# A Web-based System for Observing and Analyzing Computer Mediated Communications

Madeth May, Sébastien George, Patrick Prévôt
*ICTT Laboratory, INSA Lyon, France*
{Madeth.May, Sebastien.George, Patrick.Prevot}@insa-lyon.fr

**Abstract**

*Tracking data of user's activities resulting from Computer Mediated Communication (CMC) tools (forum, chat, etc.) is often carried out in an ad-hoc manner, which either confines the reusability of data in different purposes or makes data exploitation difficult. Our research works are biased toward methodological challenges involved in designing and developing a generic system for tracking user's activities while interacting with asynchronous communication tools like discussion forums. We present in this paper, an approach for building a Web-based system for observing and analyzing user activity on any type of discussion forums.*

## 1. Introduction

In distance learning situations, researchers are strongly interested in tracking users' activities while interacting with communication tools provided by each learning environment, and exploiting their traces. Each trace acquired from learning environments has become an important element that indicates multiple roles in contributing benefits to various actors, like researchers, developers, learners, and instructors. Thus far, a number of tracking systems and tracking data exploitation tools have been formulated and used by a large number of distance learning platforms [1], [2], [3], [4].

## 2. Identifying elements of problem

To keep tracks of user's communications in forums (i.e. display messages, post new messages, etc.), log files are used by forum developers to record every event occurring on server (where the forum is hosted). The existence of log files seems to be ignored by forum users (learners and instructors). In addition, the format of traces in log file varies from one forum to another, due to the absence of standardization and the fact that each log file depends on how it was generated. This point often causes problems for the reusability and the exploitation of existing traces. Besides, there is a lack of semantic aspects for traces stored in log files (i.e. pure text log file).

Regarding the existing tracking systems, two significant remarks have been made: (i) most systems were designed to track user's activity only on server side, the user's interaction on client side (on the remote workstation of users) is completely ignored; (ii) the activities of lurkers [5] (users who do not participate in posting messages in forum and who are not visible to other users on the forum) are not tracked down and finally, none of their traces has been recorded. Furthermore, the choices of representing the traces and the repository type to store the traces were made just to match each individual need; traces can be hard exploited independently by different exploitation tools.

To exploit the collected traces, different treatment and analysis methods have been employed. However, the major difficulties we are facing regularly are related to the effectiveness and the quality of the results returning from analysis. The analysis method used must be able to extract from traces the synthetic information that matches the specific requirement of the users. Traces as well as synthetic information extracted from the analysis might not be directly interpretable by the forum users without the assistance of trace visualization tools. Thus, friendly Graphic User Interface visualization tools should be provided to users so that they could easily interrogate traces repository by simple formal query and transform traces into different visual representations.

## 3. Research objectives and methodology

The main objective of this research is to provide a new approach for modeling traces of user's activity in all kinds of discussion forums and to response to the elements of problem we have identified above. We are working on the development of our first prototype, a

Web-based tracking system for observing and analyzing the user's activity on a contextual forum (CONFOR) [6]. The system is developed to accomplish the following objectives:

- Observe any sort of users, including lurkers, and track finely any of their activities of communications on forum on both server side and client side. Finely define the granularity of traces (compilation of tracking data on server side and tracking data on client side) and handle necessary operations such as synchronizing, structuring, updating traces, etc.
- Represent collected traces in a rich format so that they can be easily restructured and retransformed into another format of representation.
- Visualize traces in graphical representations.
- Support both instructors and students in their tasks of exploiting traces of users' activities on forum.

We have adopted one of AFNOR standards (French Association for NORmalizations, http://www.afnor.fr), which is defined for discussion forums. The choice of working with this standard allows us to (i) identify all common functionalities used for communication within forums, (ii) clearly define which user's activities are tracked within their forum communications , and (iii) build our approach of modeling traces of user's activities, without being slanted to any particular discussion forum.

Regarding the prototype development, we have designed a Web-based tracking system specifically for CONFOR and a tracking data exploitation tool, which is accessible in the long term by both instructors and students. The tests were also conducted with different use scenarios, which mainly represent communication activities of both active members and lurkers.

## 4. Modeling traces of user's activity

### 4.1. Suggested approach

The suggested approach for modeling traces of user's activities in discussion forum, presented in figure 1 can be viewed in three levels.

We name the first level as **observing and generating tracking data**. The different **observation components** are specifically designed with a number of **traces collectors**, which take care of observing and capturing the user's activities on forum. An observation component describes how an activity on forum can be performed by a user and how the trace collector generates instantaneously, the tracking data representing that activity. The raw tracking data generated by each trace collector (**TC***i* **server** and **TC***j*

**client**) will be next synchronized and transformed (**structuring and transforming tracking data**) into a structure, which conforms to a described **use model** before being stored in the trace repository.

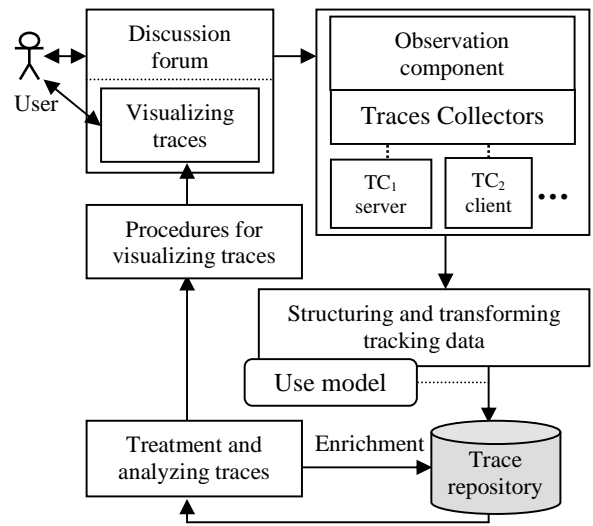

**Figure 1. Overview of suggested approach for building tracking system for discussion forums**

We have called the second level **exploiting and enriching traces**. The **trace repository** presented in this level can be a database server or a collection of files that contain the structured traces submitted from the first level. The exploitation of the traces is defined by operations such as **treatment and analyzing traces,** and a series of **procedures for visualizing traces**. The trace **enrichment** operations are for restructuring traces by adding more information to their original representations; or retransforming traces into another format, which can then be used in other exploitations tools.

At the third level, we have different Graphic User Interface modules used for **visualizing traces** (i.e. synthetic information provided by the second level). The visualization is also feasible in different visual forms, according to the parameters indicated within the information to be visualized, and independently the visualization tools.

### 4.2. Use model for user's activity on forum

A use model is used to describe the way users employ forum functionalities to perform communication activities. It allows us to (i) define the context of user's activities in forum on which the observation must be carried out, and (ii) identify every set of observable objects on forums. An **observable object** can be simply presented by a couple of the objects of interaction and its associated events. The **Objects of interaction** are present to the users within their discussion activities on forum. Each interaction between user and one of the presented objects of interaction creates an **event**, which is instantaneously captured by the trace collector.

A use model of an activity called «*Post a new message*» on forum, will be described by the objects of interaction such as the *hypertexts*, the *buttons*, the *images*, the *form for posting a new message*, etc., by which the users employ to accomplish the activity «*Post a new message*» and by their associated events such as *click*, *edit text*, *scroll*, etc. The identification of all the objects of interaction to be observed and their associated events allow the traces collectors to take into account of every user's interactions with those objects and to produce the traces of user's activities on forum in accordance with its defined use model.

### 4.3. Structuring traces

We define a general structure of trace by an alternate sequence of **states** and **transitions**. A **state** represents the instantaneous state of an activity on the forum. A **transition** is composed of one or many events produced by users' interactions during their activities on forum, which particularly make changes to the states of an activity from instant *Ti* to instant *Tj*.

Let us give an example of how we structure the traces of an activity «*Post a new message*». The event representing the user's interactions on their browser, such as typing message, moving scrollbar upward or downward, etc., will be captured by traces collectors on client side. The tracking data will be generated and temporarily stored on user's workstation. Such event does not make changes to the state of the current activity. But, if user clicks on the «*submit*» button, there is a transition: user is no longer at his/her current activity «*Post a new message*», but he is now at another state of another new activity, which might be «*Display posted message*» or «*Return to forum index*», etc. At each transition, the temporary tracking data, previously stored on client workstation, will be submitted to the server, synchronized with those on server, structured and stored in the trace repository.

## 5. A web-based system for tracking user's activity on contextual forum

### 5.1. Implementing observation component

The observation component is composed of different traces collectors, which will be used for observing user's activities on both client and server side. The traces collectors on server side are coded as a function with common parameters such as *user's information*, *type of the activity*, and *attributes of the activity*, which can be viewed in the trace repository. For the time being, these trace collectors are coded in PHP and assigned to handle the important functions, such as insertion, structuring, synchronization, and update of the traces.

We chose to develop the trace collector for client side in Java script language, because Java script is an executable language on user's Web browser (client side) and supported by any kind of Web browser. The two significant functions of trace collector on client side are: (i) to capture the users' interactions on their remote workstation; and (ii) to submit the generated tracking data to the server.

### 5.2. Implementing trace repository

We chose the relational database MySQL to implement the trace repository. The choice of using a relational database for storing traces of user's activities on forum has several advantages such as:
- Traces are structured in a rich format.
- Traces can be easily restructured and transformed into another format, such as TXT or XML.
- The operations for traces manipulation such as insertion, modification, etc., can be easily performed with simple SQL queries.

### 5.3. Managing and exploiting traces

The tools for managing the trace repository are only accessible by the instructors. These tools provide important functions to manipulate the trace repository through simple graphic user interfaces. The forum administrator (or the instructor) can easily add, or delete information used for structuring traces: i.e. the different categories of traces, different types of activities, etc.

The traces exploitation tools are designed to be more flexible. Some functions can be accessible by the students, particularly, those that are for visualizing traces. Furthermore, with these tools, users, such as researchers and instructors, can transform and export traces, which are originally in relational format into other formats of representation (TXT and XML for the time being).

### 5.4. Visualizing traces

The tools for visualizing traces in graphical representations allow the users to easily interrogate the traces repository, and to visualize the traces of user's activities in different graphical representations by a simple request query. We give below an example of visualizing traces of an activity «*Reading a message in forum*».

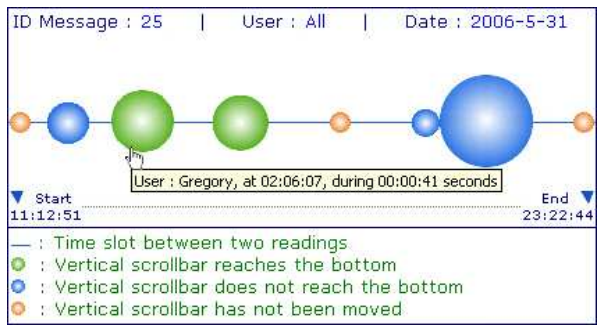

**Figure2. Visualizing traces of « Reading a message in forum » in graphical representation**

Figure 2 shows an example of visualizing traces of an activity « *Reading a message in forum* » of every users who have read the message *IDMsg = 25* on *May 31, 2006*, and between 11:12:51 and 23:22:44. The main objective of this visualization is to find if the users have displayed the message contents, if they have read it, and for how long.

Each sphere shown in figure 2 represents an activity of reading a message in forum. The diameter of the sphere is proportional to the time spent by each user for reading the displayed message. The distance between two spheres represents the time gap between two different readings. A sphere can be in one of the three colors: orange, blue, or green.

The green sphere notifies that the user read the message by having moved the vertical scrollbar downward, and to the bottom of the page (reading till the end of the message). The orange sphere expresses the fact that the user has only displayed the message contents without touching the scrollbar. The blue sphere signifies that the user has displayed the message contents, by moving the vertical scrollbar downward, but not moved to the bottom of the page (partial reading). A brief or detail information corresponding to the reading activity can be viewed by moving mouse over or clicking on each sphere.

## 6. Conclusions and future work

We have stated the potential elements of problems in tracking user's activities on computer mediated communication tools and in exploiting the traces collected by the tracking system. Our principal contribution to this challenge is an approach for modeling a common architecture for tracking systems used for observing and analyzing user's activities on CMC tools. Our developed prototype contributes to the researches in the field of tracking user's activities, particularly on discussion forums. We strongly believe that the prototype we have in hand now is an important base for us to pursue our research works.

Our future work will focus on two main objectives: (i) the evolution of the analysis methods used for treatment and analyzing traces, and (ii) the development of the tracking system, as a new extension for the various distance learning platforms.

We wish to apply in a forthcoming work the contents analysis methods in order to analyze the contents of users' communications in forum. We aim to make our system more effective in extracting from traces both quantitative and qualitative information, which reflects more behavioral, social and cognitive aspects of learners. Regarding the future tracking system, we expect to develop it in the form of a package of functions, which can be integrated in the existing distance learning platforms with limited technical skills. Besides, we are willing to carry out some experiments with students and instructors in real distance learning situations.